\documentclass[aps,amsmath,twocolumn,floatfix,superscriptaddress,prb,footinbib,10pt]{revtex4-2}

\usepackage{amsmath,amssymb,bm,mathtools,mathrsfs,comment}
\usepackage{setspace}[10]
\usepackage[italicdiff]{physics}
\usepackage{siunitx}
\usepackage{graphicx}
\usepackage[usenames, dvipsnames]{color}
\usepackage{xcolor}
\usepackage{tabularray}
\PassOptionsToPackage{hyphens}{url}
\usepackage{hyperref}
\usepackage[version=3]{mhchem}

\hypersetup{
	setpagesize=false,
	bookmarksnumbered=true,%
	bookmarksopen=true,%
	colorlinks=true,%
	linkcolor=blue,
	citecolor=red,
}

\graphicspath{{fig/}{./fig/}}

\allowdisplaybreaks

\newcommand{\bk}{{\bm{k}}}
\newcommand{\br}{{\bm{r}}}

\newcommand{\bd}{{\bm{d}}}
\newcommand{\ah}{\hat{a}}
\newcommand{\bh}{\hat{b}}

\newcommand{\bj}{{\bm{J}}}
\newcommand{\hbj}{{\hat{\bm{J}}}}

\newcommand{\brj}{{\br+\bd_{j}}}

\newcommand{\tini}{{t_{\mathrm{ini}}}}

\newcommand{\tE}{\tilde{E}}

\begin{document}
\title{Two-color laser control of photocurrent and high harmonics in graphene}
\author{Minoru Kanega}
\email{m.kanega.phys@chiba-u.ac.jp}
\affiliation{Department of Physics, Chiba University, Chiba 263-8522, Japan}
\author{Masahiro Sato}
\email{sato.phys@chiba-u.ac.jp}
\affiliation{Department of Physics, Chiba University, Chiba 263-8522, Japan}

\date{\today}
\begin{abstract}
    We comprehensively investigate two-color-laser-driven photocurrent and high harmonic generation (HHG) in graphene models. 
    By numerically solving the quantum master equation, we uniformly explore a broad parameter regime including both the weak (perturbative) and intense-laser (nonperturbative) cases while considering the dissipation effects.
    We demonstrate that the HHG spectra can be drastically altered by tuning the real-space path traced by the laser electric field. 
    This controllability is explained by the dynamical symmetry argument. We also show that both the magnitude and the direction of photocurrent (zeroth order harmonics) can be controlled by varying the frequency, intensity, ellipticity, and relative phase of the two-color laser. 
    Furthermore, the nature of photocurrent is shown to be classified into shift- or injection-current types, depending on the phase of two-color laser. 
    Our findings indicate that even in centrosymmetric electron systems, photocurrent and HHG can be quantitatively controlled by adjusting various external parameters if we utilize multicolor laser with a lower spatial or temporal symmetry.
\end{abstract}
\maketitle

\section{Introduction}\label{Sec:Intro}
Laser-induced nonequilibrium phenomena,
such as
photovoltaic effects~\cite{vonBaltz1981Theory, Sipe2000Secondorder, Cook2017Design, Young2012First, Tan2016Shift, Morimoto2016Topological, Tokura2018Nonreciprocal, Ishizuka2019Rectification, Sturman1992Photovoltaic, Watanabe2021Chiral, Dai2023Recent, Ishizuka2024Peltier},
laser-driven high-harmonic generation (HHG)~\cite{Ghimire2019Highharmonic, Yue2022Introduction, Goulielmakis2022High, Li2023High, Bhattacharya2023Strong,Hirori2024HighOrder},
and Floquet engineering~\cite{Eckardt2015Highfrequency, Mikami2016BrillouinWigner, Mori2016Rigorous, Kuwahara2016Floquet, Eckardt2017Colloquium, Oka2019Floquet, Sato2021Floquet},
are among the hottest research fields in modern condensed-matter physics.
The HHG and the photovoltaic effect, which may be viewed as photocurrent generation or zeroth-order harmonics, have attracted interest from both fundamental and applied scientific viewpoints.
For instance, the photovoltaic effect is deeply associated with solar power generation and optical sensing~\cite{Grinberg2013Perovskite, Nie2015Highefficiency, Tan2016Shift}.

It is widely known that the photovoltaic effect and even-order HHGs are prohibited by the inversion symmetry in solid electron systems if the electric dipole transition is dominant in the photoinduced dynamics.
Thus far, the research has primarily focused on noncentrosymmetric materials,
such as $p$--$n$ junctions~\cite{Sze2006Physics},
perovskite ferroelectrics~\cite{Li2021Photoferroelectric},
and Weyl semimetals~\cite{Wu2017Giant, Patankar2018Resonanceenhanced, Osterhoudt2019Colossal, Sirica2019Tracking}.

\begin{figure}[tb]
    \centering
    \includegraphics[width=\linewidth]{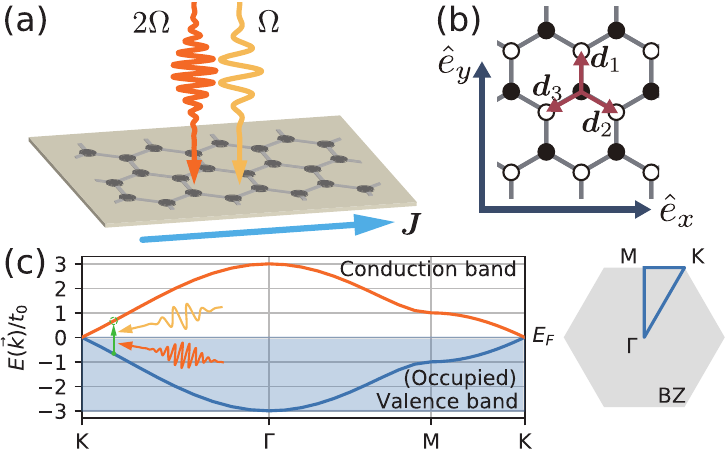}
    \caption{
        (a) Image of the photocurrent generation by a combination of two lasers with different frequencies ($\Omega$ and $2\Omega$), which generates a DC photocurrent $\bm J$.
        (b) Top view of the graphene lattice structure and the lattice vectors $\bm{d}_1$, $\bm{d}_2$, and $\bm{d}_3$. 
        Black (white) circles are A (B) sublattice sites. 
        (c) Energy band $E(\bm k)$ of graphene with Fermi energy $E_F=0$ and a photoinduced electron transition.
        The locations of $\Gamma$, $K$, and $M$ points in Brillouin zone (BZ) are depicted in the right panel.
    }
    \label{fig:figure1}
\end{figure}

As a complementary approach to the use of noncentrosymmetric materials,
attention has been drawn to using multiple external fields (multiple drive) to control optical responses or symmetries in centrosymmetric materials.
This approach can potentially extract the desired responses regardless of the material's symmetry.
Typical multiple-drive techniques include (i) application of both a DC electric field and a one-color laser and (ii) the use of two-color laser pulse, which is the main focus of this paper [Fig.~\ref{fig:figure1}(a)].

The current-induced HHG based on the first method (i) has been theoretically investigated~\cite{Khurgin1995Current,Wu2012QuantumEnhanced,Cheng2014DC,Takasan2021Currentinduced,Gao2021Currentinduced,Kanega2024Highharmonic},
and the several experiments have been reported in materials such as
Si~\cite{Aktsipetrov2009DCinduced},
GaAs~\cite{Ruzicka2012SecondHarmonic},
graphene~\cite{Bykov2012Second, An2013Enhanced},
and superconducting \ce{NbN}~\cite{Nakamura2020Nonreciprocal}.
On the other hand,
the experimental techniques for two-color laser pulses have been gradually developed~\cite{Eichmann1995Polarizationdependent,Hache1997Observation,Sun2010Coherent,Fleischer2014Spin,Bas2015Coherent,Kfir2015Generation,Baykusheva2016Bicircular,Kfir2016Inline,Kerbstadt2017Ultrashort,Kerbstadt2019Odd,Ogawa2024Programmable,Mitra2024Lightwavecontrolled,Tyulnev2024Valleytronics}.
The theoretical study of two-color-laser-driven HHG, photovoltaic effects, and Floquet engineering have also progressed in conducting electron systems~\cite{Atanasov1996Coherent,Bhat2000Optically,Rioux2012Optical,Muniz2014Coherent,Nag2019Dynamical,Kang2020Topological,Mrudul2021Lightinduced,Neufeld2021LightDriven,Rana2022Generation,Ikeda2022Floquet,Trevisan2022Bicircular,Zheng2022Optical,Avetissian2022Efficient,Ikeda2023Photocurrent,Wang2023Topological,Ikeda2024Controllable,Arakawa2024Lightinduced}, magnetic insulators~\cite{Udono2024Triplons} and molecules~\cite{Kato2024Description},
but many issues have still remained, especially, in the nonperturbative (strong-laser) regimes.

In this paper, based on the above background,
we theoretically explore the photovoltaic effects and HHG in two-color-laser-irradiated Dirac electrons in graphene~\cite{CastroNeto2009electronic, Aoki2014Physics} from weak (perturbative) to strong laser (nonperturbative) regime [see Fig.~\ref{fig:figure1}(a)].
Using numerical quantum master equation approach~\cite{Gorini1976Completely, Lindblad1976generators, Breuer2007Theory, Ikeda2019Highharmonic, Kanega2021Linear, Kanega2024Highharmonic, Tanaka2024Theory},
we quantitatively estimate the photocurrent and HHG spectra while considering the inevitable dissipation effects.
An important advantage of our numerical method is to accurately compute the HHG spectra and the photocurrent in a broad parameter space including both perturbative (weak-laser) and nonperturbative (strong-laser) regimes. 
In addition, the approach enables us to directly treat finite-length laser pulses, whose treatment is difficult within the analytic perturbative methods such as nonlinear response theories.
We demonstrate that the HHG spectra significantly depend on the real-space path traced by the laser-electric field, and that this dependence can be explained by the dynamical symmetry argument~\cite{Alon1998Selection,Liu2016Selection,Morimoto2017Floquet,Neufeld2019Floquet,Ikeda2019Highharmonic,Chinzei2020Time,Ikeda2020Highorder,Kanega2021Linear,Kanega2024Highharmonic,Udono2024Triplons}.
We also show that multiple resonant points for photocurrents appear in the nonperturbative regime,
and that the photocurrent direction can be controlled by tuning two-color laser parameters such as intensity, ellipticity, and relative phase.
By systematically exploring a broad range of these parameters, we identify many resonance behaviors and photocurrent direction flips.
We elucidate the detailed parameter dependencies and discuss the qualitative physical picture behind the phenomena.
Our findings indicate a usefulness of multicolor laser fields for generating highly controlled photocurrents and HHGs in a broad class of materials.

The remainder of the paper is organized as follows.
In Sec.~\ref{Sec:Model},
we introduce a tight-binding model for graphene coupled to a laser field and the quantum master equation for tracking the time evolution of the density matrix.
The main numerical results based on this master equation are given in Secs.~\ref{Sec:HHG}--\ref{sec:ShiftInjection}.
In Sec.~\ref{Sec:HHG}, we focus on the HHG spectrum for three types of two-color-laser configuration, and show how the HHG spectra depend on the spatial symmetry of the laser electric field path.
In Sec.~\ref{sec:frequency},
we investigate the effects of laser frequency and intensity on the photocurrent, focusing on the resonance conditions and the power-law behavior of the photocurrent.
Section~\ref{sec:EllipticityAndConductivity} reveals the ellipticity dependence of the photocurrent, including an estimate of the nonlinear optical conductivity.
In Sec.~\ref{sec:phase}, we address the relative phase dependence of the photocurrent, examining both the clean limit and the slightly dirty case.
In the clean limit, we find that the photocurrent, especially its direction,  is associated with a characteristic function of the laser electric field.
Section~\ref{sec:ShiftInjection} focuses on the relaxation-time dependence of the photocurrent to classify it into shift and injection-current types.
We demonstrate that the phase of the two-color laser field determines which type of photocurrent dominates.
Finally, in Sec.~\ref{sec:conclusions}, we summarize our findings and offer concluding remarks.
Some theoretical details related to dynamical symmetry and supplemental numerical results are discussed in Appendices.

\section{Model and Method}\label{Sec:Model}
We focus on single-layered graphene~\cite{CastroNeto2009electronic, Aoki2014Physics} with A and B sublattices [black and white circles in Fig.~\ref{fig:figure1}(b)].
The tight-binding Hamiltonian is given by
\begin{align}
    \hat{H}_0 = -t_0\sum_{\bm{r}}\sum_{j=1,2,3}(\bh_\brj^\dag\ah_\br + \ah_\br^\dag\bh_\brj),
\end{align}
where $\bd_j$, the position vectors pointing to the three nearest-neighbor sites from an A-sublattice site, are given by
$\bd_1 = a\qty(0, 1)$, $\bd_2 = a(\cos(\frac{\pi}{6}), -\sin(\frac{\pi}{6}))$, and $\bd_3 = a(-\cos(\frac{\pi}{6}), -\sin(\frac{\pi}{6}))$,
with $a$ being the lattice constant.
The vector $\br$ represents each position of sublattice A.
The electron's annihilation and creation operators on an A (B) sublattice site $\br$
are respectively $\hat a_\br$ and $\hat a_\br^\dag$
($\hat b_\br$ and $\hat b_\br^\dag)$.
They satisfy the anticommutation relations
$\acomm{\hat{c}^\dag_\br}{\hat{c}'_{\br'}}=\delta_{c,c'}\delta_{\br,\br'}$ with $c \in \{a,b\}$.
The each term of $\hat{H}_0$ describes the nearest-neighboring hopping between different sublattice sites with transfer integral $t_0$.
For graphene, $t_0$ is estimated as $t_0=\SI{2.7}{eV}$~\cite{Reich2002Tightbinding, CastroNeto2009electronic}.
Throughout the paper, we assume that the ground state is half-filled, electrons are spinless, and the Fermi energy $E_{\mathrm{F}}=0$.

To discuss the laser-driven dynamics, we adopt the Peierls phase formalism.
The time-dependent Hamiltonian for the laser-driven graphene is given by $\hat{H}(t) = -\sum_{\bm{r},j}(t_{\br,\brj}(t)\bh_\brj^\dag\ah_\br + t_{\brj,\br}(t)\ah_\br^\dag\bh_\brj)$,
where $t_{\br,\brj}(t)$ is the hopping amplitude given by the Peierls substitution, 
\begin{align}
t_{\br,\brj}(t) = t_0\exp(-i\frac{e}{\hbar}\int_\br^\brj \bm{A}(t)\cdot d\br).
\label{eq:PeierlsSub}
\end{align}
Here, $e$ is the elementary charge, $\bm{A}(t)$ is the vector potential of the laser,
and we set $\hbar=1$ throughout the paper.
This velocity gauge formalism is convenient for describing laser-driven dynamics because of its simplicity and computational efficiency, as discussed below.

We focus on a two-color laser pulse
with angular frequencies $\Omega_{1}$ and $\Omega_2$.
Hereafter, we simply refer to the angular frequency as frequency unless otherwise noted.
The vector potential $A_i(t)$ of a laser pulse with frequency $\Omega_{i}$ ($i=1,2$) is defined as
\begin{equation}
    \bm{A}_i(t)=\frac{E_i}{\Omega_i\sqrt{1+\epsilon_i^2}}f_{\mathrm{env}}(t)\mqty(\cos(\Omega_i t+\phi_i)\\\epsilon_i\sin(\Omega_i t+\phi_i)\\0),
    \label{eq:VectorPotential}
\end{equation}
where $E_i$ is the strength of the AC electric field of the pulse, $\phi_i$ is the initial phase, and $f_{\mathrm{env}}(t)$ is a Gaussian envelope function $f_{\mathrm{env}}(t)=\exp[-2(\ln 2)(t^2/t^2_{\mathrm{FWHM}})]$ with full-width at half-maximum $t_{\mathrm{FWHM}}$.
To fix the pulse width,
we adopt the five-cycle period of laser at $\Omega=0.1t_0$ as the standard of $t_{\mathrm{FWHM}}$.
The dimensionless AC field strength can be defined as $\tilde E_i = eE_i a/t_0$:
For instance, $\tilde E_i=0.01$ corresponds to $E_i\sim\SI{1.1}{MV/cm}$ in graphene.
The laser ellipticity $\epsilon_i$ denotes the degree of laser polarization: $\epsilon_i=0$ means a linear polarization (LPL) along the $x$ axis, while $\epsilon_i=\pm1$ corresponds to circular one (CPL).
The vector potential for a two-color laser is given by $\bm A(t) = \bm A_1(t) + \bm A_2(t)$.
The Dirac band persists in the energy range $|E|\alt t_0$ as shown in Fig.~\ref{fig:figure1}(c) and we will focus on the laser-driven dynamics in such a Dirac electron system. 
Therefore, the frequency $\Omega_i$ is chosen to be less than $t_0$ throughout this paper.

We compute the time evolution of the density matrix to explore the laser-driven dynamics, taking dissipation effects into account.
For this purpose, it is very important to consider the Fourier (i.e., wave-vector $\bk$) space representation of the Hamiltonian, which is given by
\begin{align}
    \hat{H}(t)=\sum_\bk\hat{H}_\bk(t)=\sum_\bk\bm{C}^\dag_\bk M(\bk+e\bm{A}(t))\bm{C}_\bk.
    \label{eq:Hamiltonina_kspace}
\end{align}
Here, $\bm{C}_\bk=\mqty(\tilde{a}_\bk & \tilde{b}_\bk)^\top$, $(\tilde{a}_\bk,\tilde{b}_\bk)$ is the Fourier transform of $(\hat{a}_\br,\hat{b}_\br)$, and $M(\bk)$ is a $2\times2$ matrix.
The time-dependent Hamiltonian is $\bk$-diagonal,
and we can thus independently solve the time evolution for each wave vector $\bk$ subspace under the assumption that dissipation effects at $\bk$ and $\bk'$ are simply independent of each other.

This $\bk$-diagonal property is a key advantage of the velocity gauge in master equation simulations.
Another commonly used gauge, the length gauge, can be obtained from the velocity gauge via a unitary transformation.
In the length gauge, the Hamiltonian consists of a static hopping term with coefficient $t_0$ and an electric potential term $\hat H_{\mathrm l.g.}(t) = -\bm E(t)\cdot\br\hat n_\br$, where $\hat n_\br$ is the number operator at site $\br$, and $\bm E(t)=-\pdv{\bm A(t)}{t}$ is the laser electric field.
Because of the $\hat H_{\mathrm l.g.}(t)$ term, the time evolution of the density matrix in the length gauge is no longer $\bk$-diagonal, resulting in a larger numerical cost compared with the velocity gauge. 
Moreover, $\hat H_{\mathrm l.g.}(t)$ breaks the spatial periodicity and thus the wave vector $\bm k$ is no longer a good quantum number. 
An important point is that physical observables are gauge invariant if one exactly computes them. 
Hence, we will use the velocity gauge throughout the paper to numerically solve the master equation. 
We note that for perturbation theories for external electric fields, some studies suggest that the length gauge may be more suitable than the velocity gauge~\cite{Aversa1995Nonlinear, Sipe2000Secondorder}.

We thereby introduce the $\bk$-decomposed master equation of the Gorini-Kossakowski-Sudarshan-Lindblad (GKSL) form~\cite{Gorini1976Completely, Lindblad1976generators, Breuer2007Theory, Ikeda2019Highharmonic, Kanega2021Linear, Kanega2024Highharmonic, Tanaka2024Theory}
\begin{align}
    \dv{\hat{\rho}_\bk(t)}{t} = -i\comm{\hat{H}_\bk(t)}{\hat{\rho}_\bk(t)} + \hat{\mathcal{D}}[\hat{\rho}_\bk(t)],
    \label{eq:quantum-master-eqre}
\end{align}
where $\hat{\rho}_\bk(t)$ is the density matrix in the $\bk$ space, $\hat{\mathcal{D}}[\hat{\rho}]=\gamma(\hat{L}_\bk\hat{\rho}\hat{L}_\bk^\dag-\frac{1}{2}\acomm{\hat{L}_\bk^\dag\hat{L}_\bk}{\hat{\rho}})$, and $\hat{L}_\bk$ are the jump operator describing a dissipation process.
The first and second terms of Eq.~(\ref{eq:quantum-master-eqre}) describe the unitary and dissipative time evolutions, respectively.
The phenomenological relaxation rate $\gamma$ represents the typical relaxation time of system $\tau\sim 1/\gamma$.
We set $\gamma/t_0$ to $10^{-2}$--$10^{-1}$, corresponding to $\tau\sim 2.4$--\SI{24}{fs}.
The initial state is set to the half-filled ground state $\hat{\rho}_\bk(\tini)=\dyad{g_\bk}{g_\bk}$ at the initial time $t=\tini$,
where $\ket{g_\bk}$ and $\ket{e_\bk}$ are the valence and conduction band occupied states at $\bk$, respectively.
We define the jump operator as $\hat{L}_\bk=\dyad{g_\bk}{e_\bk}$,
which induces an interband electron transition from the conduction to the valence band.
This jump operator satisfies the detailed balance condition at zero temperature, on which we focus in this paper.
One can generally obtain any physical observable from the numerical solution of Eq.~(\ref{eq:quantum-master-eqre}).

To study the nonlinear optical response,
we consider the electric current
\begin{align}
    \hbj(t) = \pdv{\hat{H}(t)}{\bm{A}(t)} \eqqcolon\sum_{\bk}\hbj_\bk(t),
\end{align}
and its expectation value per unit cell,
$\bj(t) = \frac{1}{N}\sum_\bk \bm{J}_\bk(t)=\frac{1}{N}\sum_\bk\expval{\hbj_\bk(t)}_t$, as observable of interest.
Here, the bracket denotes $\expval{\cdots}_t={\rm Tr}[\hat\rho(t)\cdots]$.

In the present work, we take a total of $N=160\times160$ points in an equally spaced fashion in the full Brillouin zone to numerically solve the GKSL equation of Eq.~\eqref{eq:quantum-master-eqre}.
We have verified that our numerically estimated quantities are sufficiently converged to those of the thermodynamic limit in this $\bm k$-mesh size (see Appendix~\ref{app:size}).

\section{HHG Spectra}\label{Sec:HHG}
\begin{figure}[tb]
    \centering
    \includegraphics[width=\linewidth]{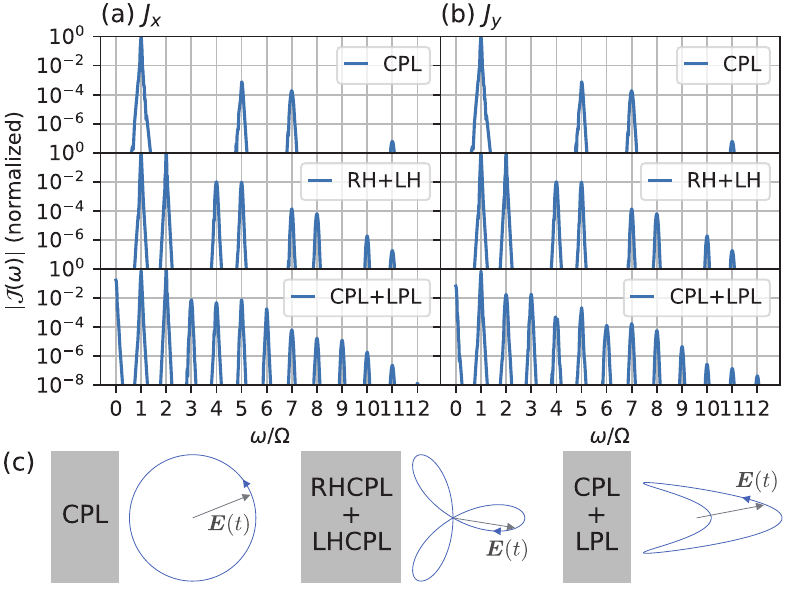}
    \caption{
        HHG spectra along the (a) $x$ and (b) $y$ axes in graphene driven by a two-color laser pulse with frequencies $\Omega_1=0.4t_0$ and $\Omega_2=0.8t_0$.
        Each panel shows the comparative plots for different laser configurations, which are
        (top row) a circularly polarized ($\epsilon_1=1$) single-laser pulse (CPL) with $\tE_1=0.1$,
        (middle row) two-laser pulses of right- and left-handed CPLs ($\epsilon_1=1$ and $\epsilon_2=-1$) with $\tE_1=\tE_2=0.05$,
        and (bottom row) two-laser pulses of CPL and a linearly polarized ($\epsilon_2=0$) pulse (LPL) with $\tE_1=\tE_2=0.05$.
        (c) Real-space paths of the three laser electric fields in the $(E_x,E_y)$ plane. 
        In all panels, $\gamma=0.1t_0$ and the phases $\phi_{1,2}=0$.
    }
    \label{fig:figure2}
\end{figure}
First, we numerically study the HHG spectra induced by two-color-laser irradiation.
We choose $\phi_{1,2}=0$ for simplicity.
Figure~\ref{fig:figure2} compares the computed HHG spectra, the Fourier transform of the current $\bm{\mathcal J}(\omega)=\int dt e^{i\omega t}\bm J(t)$, induced by three typical shapes of two-color-laser configurations.
Panel (a) [(b)] shows the $x$ [$y$] component $\mathcal J_x(\omega)$ [$\mathcal J_y(\omega)$].
Each panel depicts the results for three types of lasers:
a single (one-color) CPL [left in panel (c)], a two-color laser of right- and left-handed CPLs [middle in panel (c)], and a combination of CPL and LPL [right in panel (c)].
From these figures, it is evident that the spectra vary significantly depending on the spatial pattern of laser.
We stress that the numerical GKSL equation approach enables us to accurately estimate pulse-driven high-order harmonics without any artificial procedure such as use of window function. 
This estimation cannot be performed by perturbative methods or the numerical approach using the Schr\"odinger equation.

In the case of a single CPL, even-order harmonics and $3n$-order harmonics ($n\in\mathbb{Z}$) disappear.
When combining right- and left-handed CPLs, we find the $3n$-order harmonics disappear.
Furthermore, all-order responses appear in the combination of CPL and LPL.
Such selection rules of the appearance or absence of $j$th order harmonics are often explained by the argument based on dynamical symmetry~\cite{Alon1998Selection,Liu2016Selection,Morimoto2017Floquet,Neufeld2019Floquet,Ikeda2019Highharmonic,Chinzei2020Time,Ikeda2020Highorder,Kanega2021Linear,Kanega2024Highharmonic,Udono2024Triplons} in various systems, including atomic gases, solid-state conducting electrons, and magnetic insulators.
However, for two-color-laser-driven microscopic models, their dynamical symmetries and selection rules have not been well discussed in previous studies~\cite{Mrudul2021Lightinduced, Rana2022Generation}. 
We therefore show their proof in Appendix~\ref{app:dynamical-symmetry}.

When we apply a single CPL, the system possesses two types of dynamical symmetries associated with the two- and threefold rotational symmetries.
In the case of combining right- and left-handed CPLs, the dynamical symmetry related to the twofold rotation is broken, but the other dynamical symmetry remains, making only the $3n$-order harmonics disappear.
For the final case of CPL and LPL, both dynamical symmetries are broken, leading to all-order responses.

Our numerical result in Fig.~\ref{fig:figure2} and the above selection rule also demonstrate that zeroth-order responses, i.e., DC photocurrent generation, occur only when applied light breaks both dynamical symmetries.
Subsequent analyses of photocurrent generation will focus particularly on the two-color laser breaking these dynamical symmetries and we will assume $\tE_1=\tE_2=\tE$ for simplicity.

\section{Frequency and intensity dependence of photocurrent}\label{sec:frequency}

\subsection{Resonant conditions and intensity dependence}\label{subsec:resonance}

\begin{figure}[tb]
    \centering
    \includegraphics[width=\linewidth]{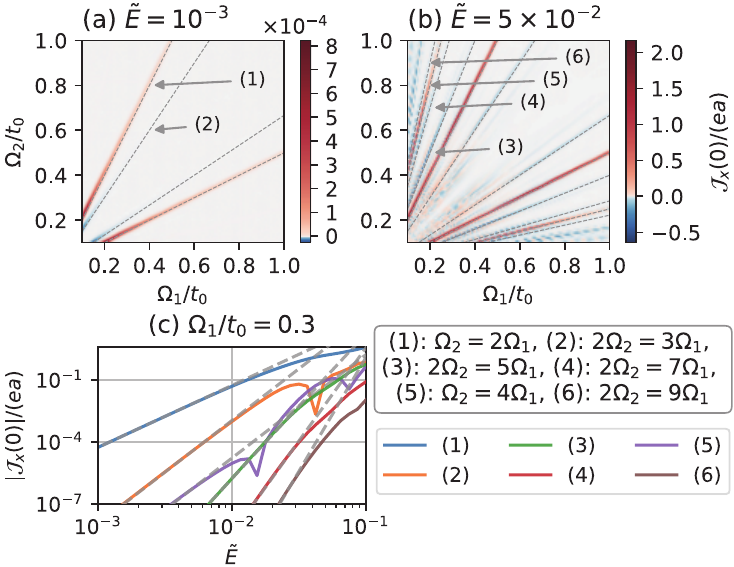}
    \caption{
    $(\Omega_1, \Omega_2)$ dependence of photocurrent $\mathcal J_x(0)$ in two-color-laser-driven graphene in (a) the weak laser ($\tE=10^{-3}$) and (b) the strong laser ($\tE=5\times 10^{-2}$) cases.
    We choose the combination of two LPLs whose polarization is parallel to the $x$ axis.
    Symbols (1)--(6) and gray dashed lines denote the resonance conditions: (1) $\Omega_2=2\Omega_1$, (2) $2\Omega_2=3\Omega_1$, (3) $2\Omega_2=5\Omega_1$, (4) $2\Omega_2=7\Omega_1$, (5) $\Omega_2=4\Omega_1$, and (6) $2\Omega_2=9\Omega_1$.
    (c) $\tE$ dependence of photocurrent $\mathcal J_x(0)$ at $\Omega_1=0.3t_0$ under the conditions of (1)--(6) in panels (a) and (b).
    In all panels, $\gamma=0.05t_0$ and $\phi_{1,2}=0$.
    }
    \label{fig:figure3}
\end{figure}
From now on, we focus on photocurrent generation (i.e., the zeroth-order harmonics). 
We discuss its dependence on laser frequency, polarization, and intensity.

First, we examine the frequency dependence.
Figure~\ref{fig:figure3} shows the photocurrent intensity generated by two-color laser as a function of laser frequencies.
Panel (a) [(b)] is the result of the weak [strong] laser, i.e., the [non] perturbative regime.
In the weak-laser case, one finds two resonance conditions (1) $\Omega_2=2\Omega_1$ and (2) $2\Omega_2=3\Omega_1$, where the photocurrent is resonantly created.
In contrast, for the strong laser, additional resonance peaks appear at (3) $2\Omega_2=5\Omega_1$, (4) $2\Omega_2=7\Omega_1$, (5) $\Omega_2=4\Omega_1$, and (6) $2\Omega_2=9\Omega_1$.

These resonance conditions can be understood by counting the photon number in the laser-driven electron excitation.
For the condition (1), photocurrent is generated through a third-order perturbative process, where two photons with frequency $\Omega_1$ are absorbed and one photon with frequency $\Omega_2$ is emitted (or its inverse process).
Hereafter, a process where $n$ photons with frequency $\Omega_1$ are absorbed and $m$ photons with frequency $\Omega_2$ are emitted will be referred to as an $(n,m)$ process.
The other resonance conditions (2)--(6) respectively correspond to
the $(3,2)$ process (fifth-order perturbation), the $(5,2)$ process (seventh-order perturbation), the $(7,2)$ process (ninth-order perturbation), the $(4,1)$ process (fifth-order perturbation), and the $(9,2)$ process (eleventh-order perturbation).

The laser intensity dependence shown in Fig.~\ref{fig:figure3}(c) supports the above argument. 
Colored lines correspond to the photocurrents at the resonance conditions (1)--(6).
The gray dashed lines are the fitting curves proportional to the cubic, fifth, seventh, ninth, fifth, and eleventh powers of the AC electric field, respectively.
In the weak-laser regime, the resonant photocurrent increases together with the gray dashed lines, indicating that the above perturbative interpretation is valid.
In the strong-laser regime, on the other hand, deviations from the fitting curves are observed.
In conditions (2) and (5), we find sharp drops of the photocurrent intensity [the peak points of curves (2) and (5) in Fig.~\ref{fig:figure3}(c)], which correspond to sign reversals of the photocurrent.
This characteristic behavior appears only in the nonperturbative (strong laser) regime.

\subsection{Power law as a function of laser frequency}\label{subsec:power-law}
\begin{figure}[tb]
    \centering
    \includegraphics[width=0.9\linewidth]{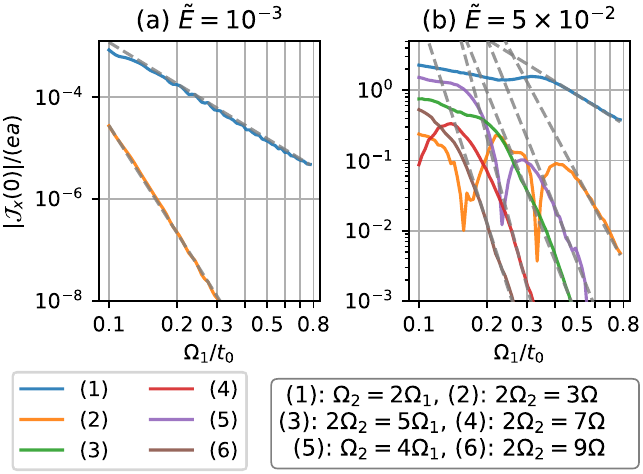}
    \caption{
        $\Omega_1$ dependence of photocurrent $\mathcal J_x(0)$ under the conditions of (1)--(6) in Figs.~\ref{fig:figure3}(a) and \ref{fig:figure3}(b).
        Dashed fitting lines in panels (a) and (b) serve as guides for the eye:
        (a) lines for the conditions (1) and (2), respectively, are proportional to $\Omega_1^{-2.7}$ and $\Omega_1^{-7}$;
        (b) lines for the conditions (1)--(6), respectively, are proportional to $\Omega_1^{-2}$, $\Omega_1^{-6}$, $\Omega_1^{-8}$, $\Omega_1^{-12}$, $\Omega_1^{-8}$, and $\Omega_1^{-10}$.
        In all panels, $\gamma=0.05t_0$, $\epsilon_{1,2}=0$, and $\phi_{1,2}=0$.
    }
    \label{fig:figure4}
\end{figure}

In addition to the power law with respect to the AC electric field under the resonant condition, we can find another power law as a function of the laser frequency. 
We compute the two-color-laser-driven photocurrent as a function of $\Omega_1$ while keeping a resonance condition ($\Omega_2=2\Omega_1$, $2\Omega_2=3\Omega_1$, etc.).
We choose the combination of two LPLs whose polarization is parallel to the $x$ axis.
In the weak laser case of Fig.~\ref{fig:figure4}(a), the photocurrent at the condition (1) is proportional to $\Omega_1^{-2.7}$ and that at the condition (2) is to $\Omega_1^{-7}$, indicating that the photocurrent intensity increases as the laser frequency decreases.
The noninteger power law $|\mathcal J_x(0)|\propto \Omega_1^{-2.7}$ at the most fundamental resonant condition (1) can be interpreted as resulting from the interplay between intraband and interband dynamics as well as finite dissipation. 
This mechanism will be discussed in the next subsection, Sec.~\ref{subsec:inter-intra-frequency}. 
On the other hand, it is difficult to give a simple interpretation for the power law $\Omega_1^{-7}$ at the condition (2). 
This may be because more complicated processes with multiple photons occur in condition (2) rather than condition (1).

In the strong-laser case of Fig.~\ref{fig:figure4}(b), the overall trend is similar, but more complex behavior is observed because of the nonperturbative effect, i.e., higher-order photon processes.
In the high frequency $\Omega_1$ region, the photocurrent at the condition (1) is proportional to $\Omega_1^{-2}$ and that at condition (2) to $\Omega_1^{-6}$, while a deviation from these power laws is observed as $\Omega_1$ decreases.
The photocurrent in condition (2) also shows a sign reversal at a certain frequency, which corresponds to the needle points of the orange line in Fig.~\ref{fig:figure4}(b).
As we have discussed in Sec.~\ref{subsec:resonance}, the photocurrents at the conditions (3)--(6) become observable only when the laser is strong enough.
Figure~\ref{fig:figure4}(b) shows that the photocurrent at the conditions (3)--(6) are proportional to $\Omega_1^{-8}$, $\Omega_1^{-12}$, $\Omega_1^{-8}$, and $\Omega_1^{-10}$ in the high-frequency $\Omega_1$ region.
Similarly to the case of the conditions (1) and (2), certain deviations from the power laws appear as $\Omega_1$ decreases.
The photocurrent at the condition (5) also shows a sign reversal.
These results suggest that use of lower-frequency laser is effective to generate larger photocurrent in both perturbative and nonperturbative regimes.

Similarly to the case of $|{\mathcal J}_x(0)|\sim \Omega_1^{-7}$ in condition (2) in Fig.~\ref{fig:figure4}(a), it is not easy to explain the above power laws $\Omega_1^{-n}$ ($n$: an integer) on different resonant conditions in the case of strong laser. 
However, we stress that the quantum master equation approach enables us to accurately estimate the $\Omega_1$ dependence even in the nonperturbative regime.

\subsection{Intraband and interband contributions and strange power law}\label{subsec:inter-intra-frequency}

Here, we try to explain the reason why we have a noninteger power-law behavior $\mathcal J_x(0)\propto \Omega_1^{-2.7}$ at the resonant condition $\Omega_1=\Omega_2/2$ in Fig.~\ref{fig:figure4}. 
To this end, we first discuss the $\Omega_1$ dependence of the photocurrent in the clean limit (i.e., for small $\gamma$).
In this section, we focus on a weak laser intensity of $\tE=10^{-3}$ and apply two LPLs whose polarization is along the $x$ axis.

\begin{figure}[tb]
    \centering
    \includegraphics[width=0.7\linewidth]{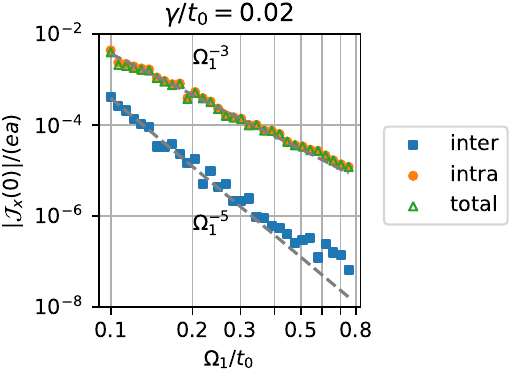}
    \caption{
    $\Omega_1$ dependence of the photocurrent along the $x$ axis under the condition of $\Omega_1=\Omega_2/2$ at $\varphi=0$.
    Blue, orange, and green dots represent the interband, intraband contributions, and the total value, respectively.
    Dashed-gray fitting lines are a guide for the eye: ${\cal J}_x(0)\propto\Omega_1^{-3}$ and ${\cal J}_x(0)\propto\Omega_1^{-5}$.
    The other parameters are $\tE=10^{-3}$, $\gamma/t_0=0.02$, and $\epsilon_{1,2}=\phi_{1,2}=0$.
    }
    \label{fig:figure5}
\end{figure}
Figure~\ref{fig:figure5} shows the $\Omega_1$ dependence of the photocurrent at $\phi_{1,2}=0$. 
We depict the interband $\bm{\mathcal J}_{\mathrm{inter}}$ (blue) and intraband $\bm{\mathcal J}_{\mathrm{intra}}$ (orange) contributions, and their total value $\bm{\mathcal J}$.
Here, the total photocurrent $\bm{\mathcal J}$ (green) is given by
\begin{equation}
    \bm{\mathcal J}(\omega) = \bm{\mathcal J}_{\mathrm{intra}}(\omega) + \bm{\mathcal J}_{\mathrm{inter}}(\omega).
\end{equation}
The intraband and interband terms are defined as
\begin{align}
    \bm{\mathcal J}_{\mathrm{intra}}(\omega) =
    \int dt \,\,\,e^{i\omega t}
    {\rm tr}[\hat \rho_{{\bm k},\mathrm{intra}}(t)\hat {\bm J}_{\bm k}(t)],\nonumber \\
    \bm{\mathcal J}_{\mathrm{inter}}(\omega)
    =\int dt \,\,\,e^{i\omega t}
    {\rm tr}[\hat \rho_{{\bm k},\mathrm{inter}}(t)\hat {\bm J}_{\bm k}(t)].
\end{align}
The density matrix is defined as $\hat \rho_{\bm k}(t)=\hat \rho_{{\bm k},\mathrm{intra}}(t)+\hat \rho_{{\bm k},\mathrm{inter}}(t)$, in which the first and second terms are, respectively, diagonal and off-diagonal components in the energy basis.
We observe that the intraband contribution is dominant, and $\mathcal J_x(0) \propto \Omega_1^{-3}$, while the interband contribution is proportional to $\Omega_1^{-5}$ in the low $\Omega_1$ region.
In the large $\Omega_1$ region, deviations from linear Dirac type dispersion begin to appear, causing a departure from the power-law behavior of the photocurrent. 
The point is that integer power laws appear in the clean enough case. 
As we will discuss later, these results are consistent with the $\gamma$ dependence shown in Fig.~\ref{fig:figure14} below.

Here, we have used a small but finite value of $\gamma=0.02t_0$ in Fig.~\ref{fig:figure5}, but we have verified the same power law even in smaller values of $\gamma$. 
Note that it is impossible to perform the numerical computation in the true clean limit of $\gamma\to 0$. 
This is because in the clean limit, the electron lifetime becomes infinite and a very long time evolution is necessary to estimate the accurate spectra of $\mathcal J(\omega)$.

\begin{figure}[tb]
    \centering
    \includegraphics[width=0.7\linewidth]{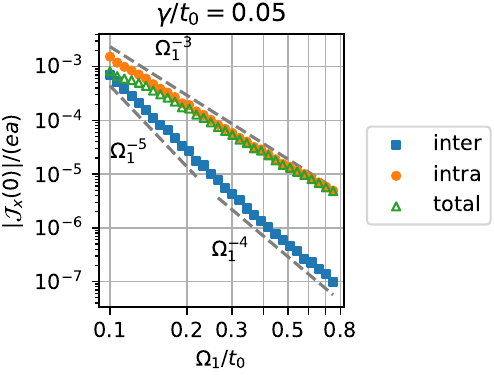}
    \caption{
    $\Omega_1$ dependence of the photocurrent $\mathcal J_x(0)$ under the condition of $\Omega_1=\Omega_2/2$.
    Blue, orange, and green dots represent the interband, intraband contributions, and the total value, respectively.
    Fitting dashed gray lines are a guide for the eye:
    $\propto\Omega_1^{-3}$, $\propto\Omega_1^{-4}$, and $\propto\Omega_1^{-5}$.
    The other parameters are set to $\tE=10^{-3}$, $\gamma/t_0=0.05$, and $\epsilon_1=\epsilon_2=\phi_{1,2}=0$.
    }
    \label{fig:figure6}
\end{figure}
Next, we return to the case of Fig.~\ref{fig:figure4}(a) with a larger $\gamma=0.05t_0$, in which higher-order terms in $\gamma$ are expected to influence the photocurrent.
Figure~\ref{fig:figure6} shows the $\Omega_1$ dependence of the interband and intraband contributions of the photocurrent under the condition (1). 
The overall behavior is similar to that in Fig.~\ref{fig:figure5}, where the intraband contribution is approximately proportional to $\Omega_1^{-3}$ and the interband one is to $\Omega_1^{-5}$.
However, with the increase in $\gamma$, particularly in the low $\Omega_1$ region,
the interband contribution significantly increases compared to Fig.~\ref{fig:figure5}.
On the other hand, the intraband contribution decreases. 
As a result, the magnitudes of the intraband and interband contributions become closer. 
Their signs are opposite and a partial cancellation occurs, and therefore the total photocurrent deviates from the simple power law $\Omega_1^{-3}$ in the low $\Omega_1$ region. 
This explains why the photocurrent is fitted by a fractional power-law function $\Omega_1^{-2.7}$ at a relatively large $\gamma$ in the small $\Omega_1$ regime.

In the high $\Omega_1$ region, where the electron band structure deviates from the Dirac-cone type,
the intraband and interband contributions also deviate from the power-law behavior, which is consistent with the trend seen in Fig.~\ref{fig:figure5}.

\subsection{Suppression of photocurrent in the nonperturbative strong-laser regime}\label{subsec:nonperturbative}

\begin{figure}[tb]
    \centering
    \includegraphics[width=\linewidth]{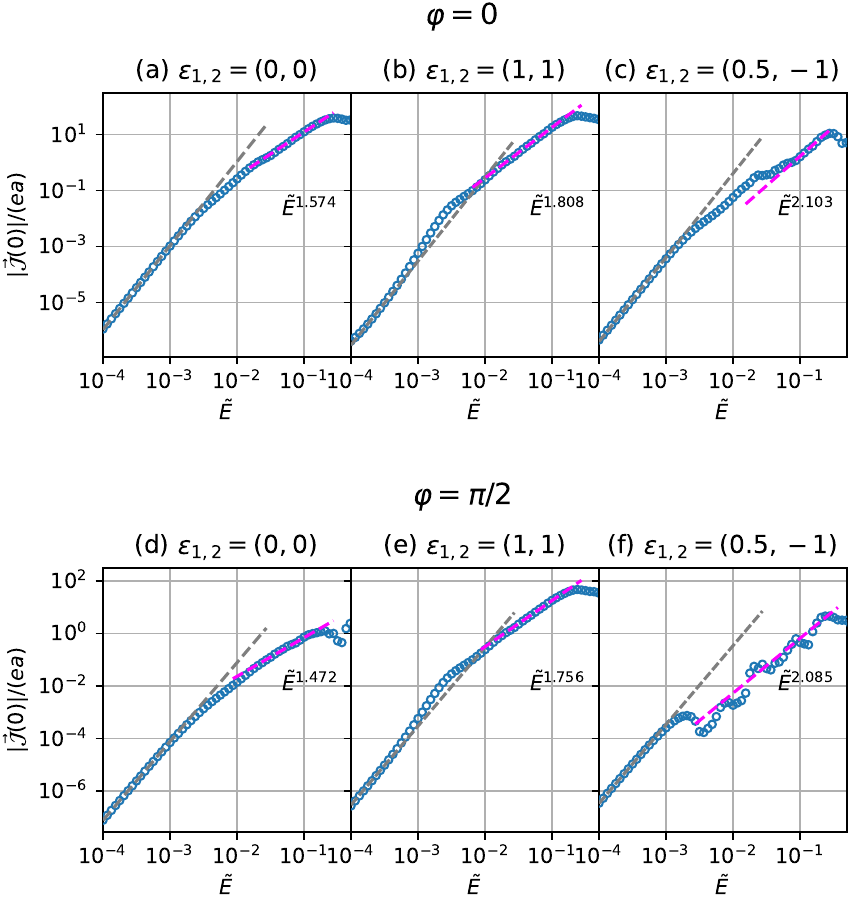}
    \caption{
        $\tE$ dependence of photocurrent $\abs{\bm{\mathcal J}(0)}$ in graphene irradiated by a two-color laser with (a)(d) $(\epsilon_1, \epsilon_2)=(0,0)$, (b)(e) $(1,1)$, and (c)(f) $(0.5,-1)$ under the resonant condition (1).
            Top (bottom) row panels are the results of $\varphi=0$ ($\varphi=\pi/2$).
            The fitting gray and red dashed lines are defined by $p_1\tE^{p_2}$, where $p_1$ and $p_2$ are fitting parameters. 
            The gray fitting lines in the weak laser regime are proportional to $\tE^{3}$ (i.e., $p_2=3$), while the fitting curves deviate from ${\tilde E}^3$ in the strong laser regime. 
            In all the numerical computations, we have set $\Omega_1=0.2t_0$, $\Omega_2=0.4t_0$, $\gamma=0.01t_0$, and $\phi_1=0$.
    }
    \label{fig:figure7}
\end{figure}

In this subsection, we consider the laser intensity dependence of the photocurrent in sufficiently clean systems irradiated by a strong two-color laser. 
Figure~\ref{fig:figure7} shows the laser intensity dependence of photocurrents in several laser-driven systems with different ellipticities and relative phases $\varphi=\phi_2-\phi_1$ under the resonant condition (1). 
The parameter $\gamma$ is set to a small value of $0.01t_0$, representing sufficiently clean systems. 
We fit the photocurrent using a polynomial function $p_1\tE^{p_2}$, where $p_1$ and $p_2$ are fitting parameters. 
In the weak laser intensity regime, the photocurrent is proportional to $\tE^3$ as expected from perturbation theory.
However, in the strong laser intensity regime ($10^{-2} \lesssim \tE \lesssim 10^{-1}$), a different power-law behavior appears to emerge. 
The value of $p_2$ varies depending on the ellipticity and phase $\varphi$, and the photocurrent follows a power law with $1 < p_2 < 3$, indicating that the growth of the photocurrent intensity is slower than the cubic power law.
In the very strong laser regime ($\tE \lesssim 10^{-1}$), the $\tilde E$ dependence of the photocurrent becomes complicated and cannot be easily fitted by a simple power-law function. 
This suppression of photocurrent growth occurs universally irrespective of values of several parameters when the laser becomes strong enough. 
A similar suppression has also been theoretically estimated in the photocurrent in one-color-laser-driven systems~\cite{Morimoto2016Topological, Kitayama2024Nonlinear}.

\section{Ellipticity dependence of photocurrent}\label{sec:EllipticityAndConductivity}

\subsection{Ellipticity dependence}\label{subsec:ellipticity}

\begin{figure}[tb]
    \centering
    \includegraphics[width=\linewidth]{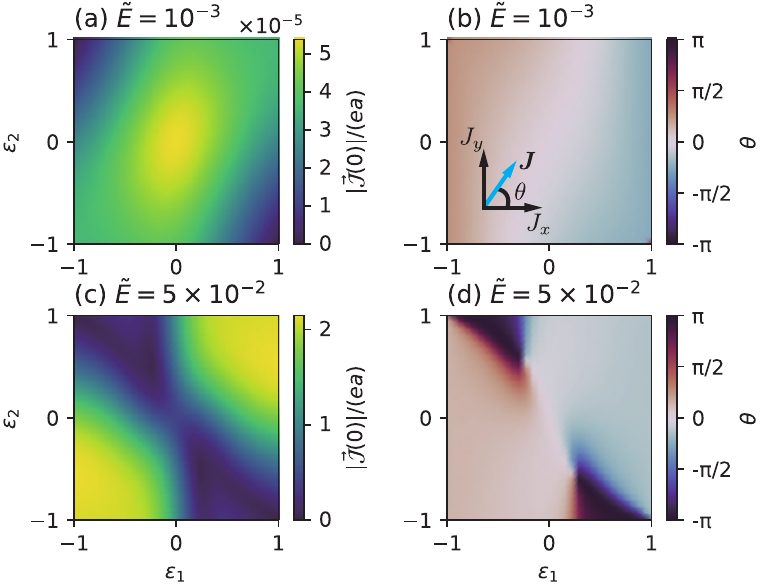}
    \caption{
    Ellipticity $\epsilon_{1,2}$ dependence of the photocurrent in two-color-laser-driven graphene in (a)(b) the weak laser $\tE=10^{-3}$ and (c)(d) the strong laser $\tE=5\times 10^{-2}$ cases.
    Panels (a) and (c) show the intensity of the photocurrent, while panels (b) and (d) depict the direction of the photocurrent on the $(x,y)$ plane.
    The angle $\theta$ is defined in the inset of panel (b).
    In all the panels (a)--(d), we have set $\Omega_1=0.2t_0$, $\Omega_2=0.4t_0$, $\gamma=0.1t_0$, and $\phi_{1,2}=0$. 
    }
    \label{fig:figure8}
\end{figure}
In this section, we study the laser ellipticity dependence of the photocurrent.
Figures~\ref{fig:figure8} depict the photocurrent intensity and its angle $\theta$ [see Fig.~\ref{fig:figure8}(b)] as a function of ellipticities $\epsilon_{1,2}$ of the two-color laser in the weak and strong laser cases.
We set the laser frequencies to satisfy the fundamental resonant condition $\Omega_2=2\Omega_1$.
The results for the weak lasers are in Figs.~\ref{fig:figure8}(a) and \ref{fig:figure8}(b).
They demonstrate that the most efficient laser for generating the largest photocurrent is given by the combination of two LPLs ($\epsilon_1=\epsilon_2=0$).
Panel (b) indicates that the corresponding angle is $\theta=0$, namely, the photocurrent is almost parallel to the positive $x$ direction.

The photocurrent behavior changes qualitatively in the strong-laser case, as shown in panels (c) and (d).
Panel (c) shows that the most efficient laser is no longer the set of two LPLs but rather the combination of two CPLs with the same rotation ($\epsilon_1=\epsilon_2=\pm1$).
Furthermore, panel (d) indicates that there are regions where the photocurrent is parallel to the negative $x$ direction ($\theta\sim \pi$), 
in contrast to the weak laser case.

\subsection{Nonlinear optical response coefficients}\label{subsec:coefficients}

To more deeply understand the $\epsilon_{1,2}$ dependence of photocurrent, we numerically estimate the nonlinear optical response coefficients $\chi_{3,5,7}$, defined by 
\begin{align}
    \mathcal J_{x}(0)=\chi_3 \tilde E^3 + \chi_5 \tilde E^5 + \chi_7 \tilde E^7+\cdots. 
\end{align}
\begin{figure}[tb]
    \centering
    \includegraphics[width=\linewidth]{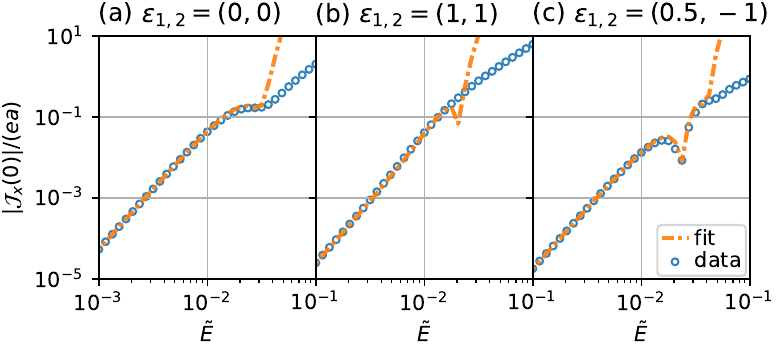}
    \caption{
        $\tE$ dependence of photocurrent along the $x$-axis under the conditions of (a) $(\epsilon_1, \epsilon_2)=(0,0)$, (b) $(1,1)$, and (c) $(0.5,-1)$.
        The fitting curves are defined by $\mathcal J_x(0)=\chi_3\tE^3+\chi_5\tE^5+\chi_7\tE^7$.
        In all the numerical computations, we set $\Omega_1=0.2t_0$, $\Omega_2=0.4t_0$, $\phi_{1,2}=0$, and $\gamma=0.1t_0$.
    }
    \label{fig:figure9}
\end{figure}
Figures~\ref{fig:figure9}(a)--\ref{fig:figure9}(c) show the laser intensity dependence for three combinations of two ellipticities $\epsilon_{1,2}$:
(a) ($\epsilon_1,\epsilon_2)=(0,0)$, (b) $(1,1)$, and (c) $(0.5,-1)$.
Perturbation theory indicates that the DC photocurrent can be expressed by an odd function of $\tilde E$; that is, any term proportional to even powers of the electric field does not appear.
The numerical data in Fig.~\ref{fig:figure9} are fitted by $\mathcal J_{x}(0)=\chi_3 \tilde E^3 + \chi_5 \tilde E^5 + \chi_7 \tilde E^7$.
We see that the photocurrent intensity varies significantly as a function of ellipticity and laser intensity.
As expected, in the weak laser intensity regime, the numerical results fit well with the leading term $\chi_3 \tilde E^3$.
In the strong laser regime, on the other hand, the simple function of $\chi_3 \tilde E^3$ no longer nicely fits the numerical results, and higher-order terms with $\chi_{5,7}$ are necessary.
In panel (a), we determine $\chi_3>0$, $\chi_5<0$, and $\chi_7>0$, while we obtain $\chi_3>0$, $\chi_5>0$, and $\chi_7<0$ in panel (b). 
The negative sign of $\chi_5$ leads to the decrease of the photocurrent in the moderate field range ($10^{-2}<\tilde E<10^{-1}$) in case (a) of $(\epsilon_1,\epsilon_2)=(0,0)$.

In panel (c), we find a needle like point at a certain AC-field intensity of $\tilde E\sim {\cal O} (10^{-2})$, 
where the photocurrent rapidly drops and sign reversal takes place.
The signs of estimated coefficients, $\chi_3>0$, $\chi_5<0$, and $\chi_7>0$ are the same as those in panel (a), but a larger value of $|\chi_5|$ appears in panel (c). 
As a result, the sign change of the photocurrent occurs as a result of the competition between $\chi_3 \tilde E^3$ and $\chi_5 \tilde E^5$.

From these results, we see that
the laser ellipticity strongly affects the magnitude of photocurrents and the nonlinear optical response functions $\chi_{3,5,7}$. 
The results also imply that one can control the photocurrent by tuning the ellipticity of the two-color laser.

\section{Phase dependence of photocurrent}\label{sec:phase}

So far, we have mainly considered the case of the phase $\phi_{1,2}=0$. 
This section focuses on the phase dependence of the photocurrent.

\subsection{Clean limit}\label{subsec:lowgamma}

\begin{figure}[tb]
    \centering
    \includegraphics[width=\linewidth]{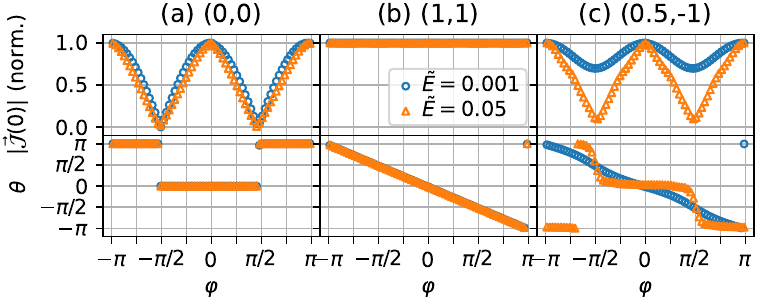}
    \caption{
        Relative phase $\varphi$ dependence of the magnitude (upper panels) and direction (lower panels) of the photocurrent in the cases of ellipticities (a) $(\epsilon_1,\epsilon_2)=(0,0)$, (b) $(1,1)$, and (c) $(0.5, -1)$. 
        We consider a sufficiently clean system with a small value of $\gamma=0.01t_0$. 
        The frequency is set at the resonance condition of $\Omega_1=\Omega_2/2=0.2t_0$, and we set $\phi_1=0$. 
        Blue and orange points are respectively the results of weak ($\tilde E=0.001$) and strong ($\tilde E=0.05$) lasers. 
        The magnitude of the photocurrent is normalized such that its maximum value takes unity.
    }
    \label{fig:figure10}
\end{figure}

In this subsection, we consider the clean limit ($\gamma$ is sufficiently small).
Figures~\ref{fig:figure10}(a)--\ref{fig:figure10}(c) show the $\varphi=\phi_2-\phi_1$ dependence of the magnitude and angle $\theta$ of the photocurrent in three different combinations of $\epsilon_{1,2}$.
These results indicate that the phases $\phi_{1,2}$ can also be used as control parameters for the photocurrent, together with tuning other parameters (laser intensity, ellipticities $\epsilon_{1,2}$, etc.).
The result for the weak laser case at a small $\gamma=0.01t_0$ [blue curves in Figs.~\ref{fig:figure10}(a)--\ref{fig:figure10}(c)] is consistent with the perturbative analysis of Ref.~\cite{Ikeda2024Controllable}, while the $\phi_{1,2}$ dependence of the photocurrent generally changes when the laser intensity increases, especially, in the case of panel (c).

\begin{figure}[tb]
    \centering
    \includegraphics[width=\linewidth]{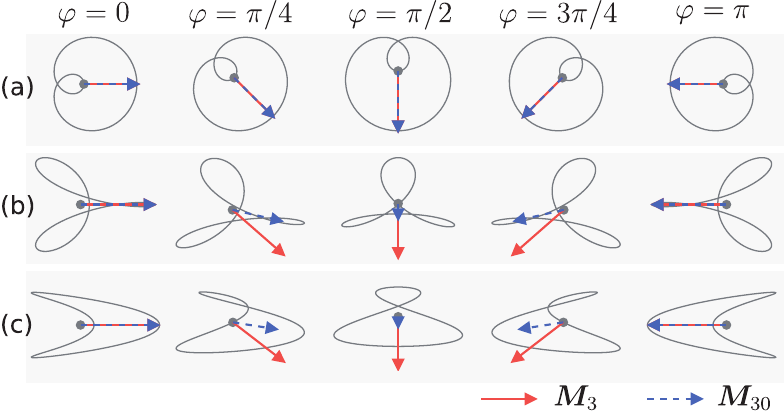}
    \caption{
        Schematic diagrams of the vectors $\bm{M}_3$ (orange solid arrow) and $\bm{M}_{30}$ (blue dashed arrow) for several combinations of laser ellipticity and relative phase.
        The relative phases are chosen as $\varphi=\{0,\pi/4,\pi/2,3\pi/4,\pi\}$.
        The ellipticity parameters $\epsilon_{1,2}$ are set to (a) $(1,1)$, (b) $(1,-0.5)$, and (c) $(1,0)$.
        Gray curves represent the real-space paths of the laser electric-fields in the $(E_x,E_y)$ plane.
        The arrows are normalized in each panel.
    }
    \label{fig:figure11}
\end{figure}

To examine in greater detail the dependence on the relative phase, we introduce the vector quantity $\bm{M}_n$ as an indicator of nature of the photocurrent. 
The vector $\bm{M}_n$ is defined as
\begin{align}
    \bm M_n = \int_0^T \dd t \abs{\bm E(t)}^{n-1} \bm E(t),
\end{align}
where $\bm E(t)$ is the laser electric field, 
$T=2\pi/\Omega_1$ is the laser period, and $n$ is an integer. 
This quantity is interpreted as the time average of a nonlinearly weighted electric field over one period.
When evaluated for $n=1$, one finds that $\bm{M}_1 = \bm{0}$. 
This is because the time average of any laser field is always zero.  
On the other hand, $\bm{M}_{n\geq2}$ is generally nonzero. 
With increasing $n$, the vector $\bm{M}_{n\geq2}$ more sharply indicates the direction to which the laser field points on average over a period $T$. 
Figure \ref{fig:figure11} illustrates the meaning of $\bm{M}_{n\geq2}$.

\begin{figure}[tb]
    \centering
    \includegraphics[width=\linewidth]{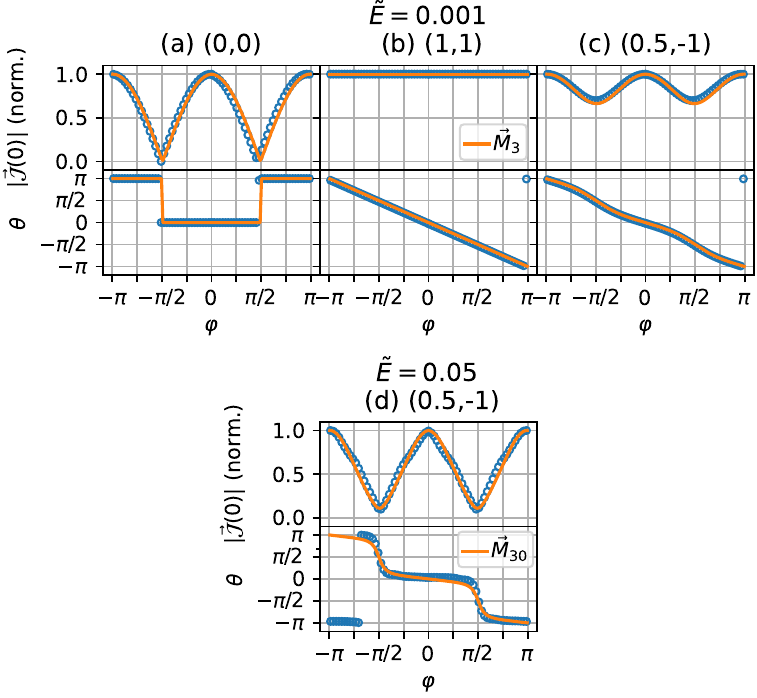}
    \caption{
        Relative phase $\varphi$ dependence of the magnitude and direction of the photocurrent $\bm{\mathcal{J}}(0)$ for the four ellipticity cases: (a) $(\epsilon_1,\epsilon_2)=(0,0)$, (b) $(1,1)$, (c) $(0.5, -1)$, and (d) $(0.5, -1)$.
        Panels (a)--(c) illustrate the weak laser case ($\tE=0.001$) and compare the photocurrent with $\bm{M}_3$.
        Panel (d) shows the strong laser case ($\tE=0.05$) and compares the photocurrent with $\bm{M}_{30}$.
        We have considered the clean limit case with a small value of $\gamma=0.01t_0$. 
        The frequency is set to a resonance condition of $\Omega_1=\Omega_2/2=0.2t_0$, and we set $\phi_1=0$.
        The magnitude of the photocurrent is normalized such that its maximum value is unity.
    }
    \label{fig:figure12}
\end{figure}

Figures~\ref{fig:figure12}(a)--\ref{fig:figure12}(c) compare the numerically computed photocurrent, corresponding to Figs.~\ref{fig:figure10}(a)--\ref{fig:figure10}(c), with $\bm{M}_3$ in the weak-laser regime.
These panels in Fig.~\ref{fig:figure12} clearly show that $\bm{M}_3$ almost perfectly reproduces the relative phase dependence of the photocurrent.
The main reason $\bm{M}_3$ coincides with the photocurrent would be that the photocurrent is proportional to $\tilde {\bm E}^3$ in the weak-laser (perturbative) regime.

We next consider the nonperturbative regime, where the laser intensity is sufficiently large. 
In this regime, higher-order perturbation terms, such as those involving the fifth and seventh powers of the electric field, contribute to the photocurrent in addition to the cubic term, making it difficult to describe the current solely with $\bm{M}_3$. 
We therefore compare $\bm{M}_n$ with the relative phase dependence of the photocurrent, varying the integer $n$. 
As a result, we find that $\bm{M}_{30}$ closely reproduces the relative phase dependence of the photocurrent in a nonperturbative case of $\tilde E=0.05$, as shown in Fig.~\ref{fig:figure12}(d). 
The choice $n=30$ does not carry special significance in itself. 
In fact, if we change the value of the AC electric field around $\tilde E=0.05$, 
the appropriate integer $n$ also changes from 30. 
The point is that the proper integer $n$ monotonically increases with the increasing AC electric field strength.

From these results, we conclude that $\bm{M}_3$ is a useful indicator for understanding the relative phase dependence of the photocurrent in the weak-laser (perturbative) regime, while $\bm{M}_n$ with larger $n$ provides a more appropriate descriptor in the strong-laser (nonperturbative) regime.

\subsection{Moderately dirty systems}\label{subsec:dirty}

\begin{figure}[tb]
    \centering
    \includegraphics[width=\linewidth]{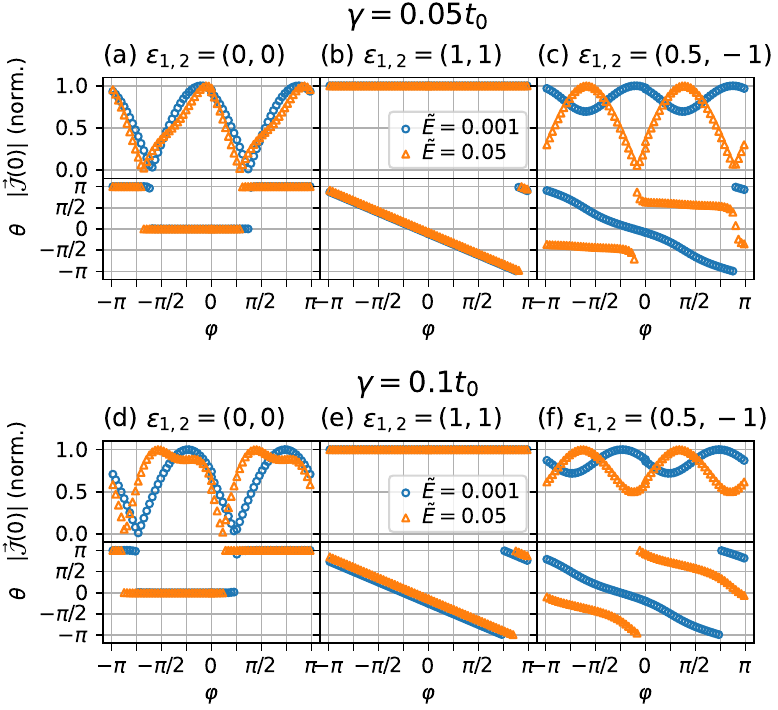}
    \caption{
        Relative phase $\varphi$ dependence of the magnitude and direction of the photocurrent in the cases of ellipticities (a)(d) $(\epsilon_1,\epsilon_2)=(0,0)$, (b)(e) $(1,1)$, and (c)(f) $(0.5, -1)$.
        The magnitude of the photocurrent is normalized so that its maximum value is unity.
        Panels (a)--(c) are the results of $\gamma=0.05t_0$, while panels (d)--(f) are those of $\gamma=0.1t_0$.
        We have set $\Omega_1=0.2t_0$, $\Omega_2=0.4t_0$, and $\phi_1=0$.
    }
    \label{fig:figure13}
\end{figure}

The last subsection focuses on the clean limit, while we comment on the case of moderate value of $\gamma$ in this subsection.
Comparing the numerical results of $\gamma=0.01t_0$ [Fig.~\ref{fig:figure10}(a)--\ref{fig:figure10}(c)], $0.05t_0$ [Fig.~\ref{fig:figure13}(a)--\ref{fig:figure13}(c)] and $0.1t_0$ [Fig.~\ref{fig:figure13}(d)--\ref{fig:figure13}(f)],
we find that the $\varphi$ dependence of the photocurrent largely changes with increasing $\gamma$. 
Figures~\ref{fig:figure10} and \ref{fig:figure13} show that photocurrents in the case of a moderately large $\gamma$ (Fig.~\ref{fig:figure13})
clearly differ from those in the clean limit (Fig.~\ref{fig:figure10}).
We also find that the magnitude and direction of the photocurrent generally change when we tune the AC-field intensity $\tilde E$. 
In particular, we find a large AC-field dependence in the cases of $(\epsilon_1,\epsilon_2)=(0.5,-1)$ in Figs.~\ref{fig:figure13}(c) and \ref{fig:figure13}(f).

\section{Shift and injection currents}\label{sec:ShiftInjection}

In this section, we discuss the relaxation-time dependence of the photocurrent in two-color-laser-driven graphene. 
In our GKSL equation formalism, the relaxation time (lifetime) of the electron is given by $\tau\propto 1/\gamma$. 
When the electron lifetime is sufficiently long $\tau \to \infty$ ($\gamma/t_0\to 0$) (i.e., the crystal is clean enough) and the laser field is sufficiently weak (i.e., the system is in a perturbative regime), we can classify the photocurrent from its $\tau$ dependence: 
If the relation $\mathcal J(0) \propto \tau^{0}$ holds, the photocurrent is referred to as a shift current, while if the photocurrent satisfies $\mathcal J(0) \propto \tau^{1}$, it is referred to as an injection current~\cite{vonBaltz1981Theory, Sipe2000Secondorder, Dai2023Recent}.

\begin{figure}[tb]
    \centering
    \includegraphics[width=\linewidth]{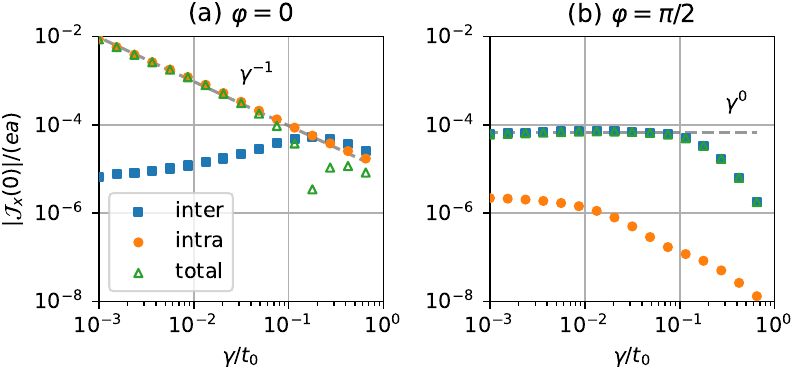}
    \caption{
    $\gamma$ dependence of the photocurrent $\mathcal J_x(0)$ under the condition of $\Omega_1=\Omega_2/2=0.2t_0$.
    Panels (a) and (b) show the results of $\varphi=0$ and $\pi/2$, respectively ($\varphi=\phi_2-\phi_1$).
    Blue, orange, and green dots represent the interband, intraband, and total photocurrents, respectively.
    Dashed gray lines of panels (a) and (b) are a guide for the eye: The fitting lines are proportional to (a) $\gamma^{-1}$ and (b) $\gamma^{0}$, respectively.
    The other parameters are set to $\tE=10^{-3}$ and $\epsilon_1=\epsilon_2=\phi_1=0$.
    }
    \label{fig:figure14}
\end{figure}

Below, we focus particularly on the condition $\Omega_2=2\Omega_1$, unless otherwise noted.
Figure~\ref{fig:figure14} shows the $\gamma$ dependence of the photocurrent generated by the two LPLs whose polarization direction is parallel to the $x$ axis.
Panels (a) and (b) respectively represent the results of (a) the relative phase $\varphi=\phi_2-\phi_1=0$ and (b) $\varphi=\pi/2$.
The generated photocurrent $\bm{\mathcal J}$ (green) is decomposed into the intraband contribution $\bm{\mathcal J}_{\mathrm{intra}}$ (orange) and the interband contribution $\bm{\mathcal J}_{\mathrm{inter}}$ (blue). 
From Fig.~\ref{fig:figure14}, we find that the relaxation time dependence of the photocurrent is significantly changed by the relative phase $\varphi$. 
First, we consider panel (a) for $\varphi=0$.
When $\gamma/t_0 \lesssim 10^{-1}$, the intraband contribution $\bm{\mathcal J}_{\mathrm{intra}}(\omega=0)$ is dominant, 
and the photocurrent is proportional to $\gamma^{-1}$.
Therefore, the main part of the photocurrent is of the injection type. 
When $\gamma/t_0 \agt 10^{-1}$, in which the system is far from the clean limit,
the interband and intraband contributions become comparable.
These two contributions have opposite signs, leading to a sign reversal of the total photocurrent.

Next, we consider panel (b) for $\varphi = \pi/2$.
In this case, when $\gamma/t_0 \lesssim 10^{-1}$, the primary contribution comes from the interband dynamics,
and the photocurrent is almost proportional to $\gamma^0$.
Therefore, we can say that the photocurrent is classified as a shift current in the case of $\varphi = \pi/2$.
When the lifetime becomes shorter like $\gamma/t_0 \agt 10^{-1}$, the photocurrent is no longer proportional to $\gamma^0$, but the interband components still remain dominant.
The results for $\varphi=0$ and $\pi/2$ in the clean limit are consistent with a recent perturbative analysis in Ref.~\cite{Ikeda2024Controllable}.

In conclusion, by adjusting the relative phase, it is possible to control the type of the photocurrent from shift to injection currents or vice versa.
We also find that depending on the strength of the dissipation and the phase $\varphi$, the ratio between intraband and interband contributions drastically changes.
The phase degree of freedom emerges in the two-color-laser case and does not in the single-color laser.
Therefore, controlling shift or injection current is a unique feature of multiple-laser-driven systems.

\section{Conclusions}\label{sec:conclusions}

In this study, we have conducted a detailed numerical investigation of the photocurrent and HHG spectra in graphene irradiated by a two-color-laser.
By numerically solving the quantum master equation, we have explored a broad parameter space. 
The advantages of the numerical method using the quantum master equation are that (i) it takes into account dissipative effects and (ii) it enables us to compute physical quantities in not only the weak-laser but also the strong-laser regimes. 
Other analytical or perturbative approaches cannot be applied to estimate higher-order harmonics and observables in the strong-laser regime.

From Sec.~\ref{Sec:HHG} onward, we present the numerical results of our study. 
In Sec.~\ref{Sec:HHG}, we first show the HHG spectra induced by three typical types of two-color laser fields. 
In particular, we confirm that the presence or absence of $n$th-order harmonics changes depending on the real-space laser-field path and demonstrate, based on a dynamical symmetry argument, why this characteristic behavior arises. 
This result clearly indicates that the HHG spectra can be controlled if we change the wave shape of the two-color laser by tuning the laser ellipticity and frequency.

In Sec.~\ref{sec:frequency}, we analyze the dependence of the photocurrent on both the laser frequency and intensity. 
In the weak-laser (perturbative) regime, we find the peaks of the photocurrent appear under two resonance conditions $\Omega_2=2\Omega_1$ and $2\Omega_2=3\Omega_1$, and the photocurrent increases in a divergent fashion as the laser frequency approaches zero ($\Omega\rightarrow0$). 
In the strong-laser regime, we observe many resonant points in addition to the above conditions. 
On the resonant points, we find nontrivial power laws of the photocurrent as a function of laser intensity or frequency. 
Moreover, in some resonant cases, we show that the magnitude of the photocurrent drastically changes and even reverses sign when we tune the laser intensity or frequency. 
To deeply understand the power law on the simplest resonant point $\Omega_2=2\Omega_1$, 
we decompose the total photocurrent into the interband and intraband contributions.

Section~\ref{sec:EllipticityAndConductivity} is devoted to the discussion about the laser ellipticity dependence of the photocurrent. 
We find that this dependence differs markedly between the weak-laser (perturbative) and strong-laser (nonperturbative) regimes, and that the optimal ellipticity for generating an efficient photocurrent changes accordingly. 
We also show that the photocurrent direction can be controlled by tuning the ellipticity and the laser intensity, and that significant directional changes arise at certain parameters in the crossover from the perturbative to the nonperturbative regime.
We show that some of these ellipticity-dependent effects can be understood by estimating the nonlinear optical response functions $\chi_{3,5,7}$ for the photocurrent.

In Sec.~\ref{sec:phase}, we analyze the dependence of the photocurrent on the relative phase $\varphi=\phi_2-\phi_1$ of a two-color laser. 
We verify that varying the relative phase can drastically change the direction of the photocurrent. 
For the weak-laser case, our results agree with the perturbative analysis in Ref.~\cite{Ikeda2024Controllable}, but as the laser intensity increases, the $\varphi$ dependence of the photocurrent generally changes, which is especially pronounced in the case shown in Fig.~\ref{fig:figure10}(c). 
Furthermore, we show that an introduced vector quantity $\bm{M}_n$, whose direction is a ``time-averaged'' direction of the laser electric field, can capture the phase dependence of the photocurrent. 
Through this analysis, we demonstrate that the direction and magnitude of the photocurrent can be inferred from the real-space patterns traced by the laser electric field. 
We also confirm that the phase dependence of the photocurrent changes significantly depending on the value of $\gamma$.

In the last section, Sec.~\ref{sec:ShiftInjection}, we investigate the relaxation-time ($\tau\sim 1/\gamma$) dependence of the photocurrent, by analyzing its interband and intraband parts.
We find that the relaxation-time dependence drastically changes by varying the relative phase $\varphi$. 
When $\varphi=0$, the intraband contribution dominates, whereas for $\varphi=\pi/2$, the interband contribution does. 
Observing the $\gamma$ dependence of the photocurrent in a clean enough regime, we find that the intraband contribution at $\varphi=0$ is of an injection current type, while the interband contribution at $\varphi=\pi/2$ is regarded as a shift current. 
On the other hand, in the regime of relatively large $\gamma$, higher-order terms in $\gamma$ lead to more complex behaviors in the photocurrent.

From the analyses in Secs.~\ref{Sec:HHG}--\ref{sec:ShiftInjection}, we have unveiled various parameter dependencies of the photocurrent and HHG spectra in graphene irradiated by two-color laser. 
In particular, the ellipticity $\epsilon_{1,2}$, relative phase $\varphi$ and relaxzation-time $\tau$ dependence of the photocurrent are nontrivial. 
Moreover, the parameter dependence in the strong-laser regime is accurately estimated by using the numerical approach based on the quantum master equation. 
This study provides effective guidance for the quantitative control of photocurrents and HHG spectra in two-color-laser-driven graphene.

\begin{acknowledgments}
    M.K. was supported by JST, the establishment of university fellowships towards the creation of science technology innovation (Grant No. JPMJFS2107), and by JST SPRING (Grant No. JPMJSP2109). 
    M.S. was supported by JSPS KAKENHI (Grants No. JP17K05513, No. JP20H01830, No. JP20H01849, No. JP25K07198, No. JP25H02112, No. JP22H05131, No. JP23H04576, No. JP25H01609 and No. JP25H01251) and JST, CREST Grant No. JPMJCR24R5.
\end{acknowledgments}


\appendix

\section{Selection rules in two-color-laser driven graphene}\label{app:dynamical-symmetry}

Here, we discuss the selection rules for HHG spectra derived from dynamical symmetries.
These symmetries exactly hold in the case of continuous wave laser irradiation, i.e., $t_{\mathrm{FWHM}}\rightarrow \infty$.
However, it is known that even in the case of laser pulse with a finite $t_{\mathrm{FWHM}}$, the selection rules can be used at least qualitatively. 
In fact, our numerical results for a pulse laser are approximately consistent with the selection rules below.

When a single CPL (circularly polarized laser) is applied to graphene, the time-dependent system exhibits dynamical symmetries related to two- and threefold rotational symmetries ($n=2,3$),
\begin{align}
    \label{eq:dynamical-symmetry1}
     & \hat U_n \hat H(t+T/n)\hat U_n^\dag = \hat H(t),            \\
     & \hat U_n \hat{\bm J}(t+T/n)\hat U_n^\dag = \mathcal R^{-1}(2\pi/n)\hat{\bm J}(t).
    \label{eq:dynamical-symmetry2}
\end{align}
Here, $\hat U_n$ is a unitary operator of a $2\pi/n$ rotation in the $(x,y)$ plane, $T=2\pi/\Omega$ is the time period of the periodically driven graphene, and
\begin{align}
    \mathcal R(\theta_{\mathrm R}) = \mqty(\cos\theta_{\mathrm R} & -\sin\theta_{\mathrm R} & 0 \\ \sin\theta_{\mathrm R} & \cos\theta_{\mathrm R} & 0 \\ 0 & 0 & 1)
\end{align}
is a $U(1)$ rotation matrix around the $z$ axis by a angle $\theta_{\mathrm R}$.
In general, if a time periodic system with a Hamiltonian $\hat H(t)=\hat H(t+T)$ satisfies a symmetry relation including a time shift, $\hat U \hat H(t+\lambda T)\hat U^\dag = \hat H(t)$  ($0<\lambda<1$) like Eq.\eqref{eq:dynamical-symmetry1}, we say that the system has a dynamical symmetry.
Equation~\eqref{eq:dynamical-symmetry2} is derived from the fact that graphene has a sixfold rotational symmetry $\hat C_6$ and the real-space path of CPL has a continuous rotational symmetry $\hat C_{\infty}$.
In Ref.~\cite{Kanega2024Highharmonic}, it is shown that the above dynamical symmetries lead to the following selection rules for the Fourier components of the currents $\bm{\mathcal J}(\omega)$:
\begin{align}
    \bm{\mathcal J}(2\ell\Omega)=\bm 0, \\
    \bm{\mathcal J}(3\ell\Omega)=\bm 0,
\end{align}
where $\ell\in\mathbb{Z}$. 
For more detail of the derivation, see Ref.~\cite{Kanega2024Highharmonic}.

Next, we consider graphene irradiated by two-color laser of two CPLs, left-handed CPL with frequency $\Omega$ and right-handed CPL with frequency $2\Omega$ (or the inverse combination).
In this setup, we obtain the following relation for the vector potential:
\begin{align}
    &\bm A(t+T/m) \notag\\
    & = \bm A_1(t+T/m) + \bm A_2(t+T/m) \notag                                      \\
                 & = \mathcal R^{-1}(2\pi/m)\bm A_1(t) + \mathcal R(4\pi/m)\bm A_2(t)\notag      \\
                 & = \mathcal R^{-1}(2\pi/m)\bm A_1(t) + \mathcal R^{-1}(2\pi(m-2)/m)\bm A_2(t),
    \label{eq:vector-potential}
\end{align}
where $m$ is an arbitrary integer.
In Eq.~(\ref{eq:vector-potential}), we have used the relation $\mathcal R(\theta_{\mathrm R}) = \mathcal R^{-1}(2\pi - \theta_{\mathrm R})$.
If Eq.~(\ref{eq:vector-potential}) can be re-expressed as $\bm A(t+T/m) = \mathcal R^{-1}(g(m))\bm A(t)$ with $g(m)$ being a function of $m$, we can lead to a dynamical symmetry like Eqs.~(\ref{eq:dynamical-symmetry1}) and (\ref{eq:dynamical-symmetry2}). 
To satisfy this condition, one may require the equality $\mathcal R^{-1}(2\pi/m)=\mathcal R^{-1}(2\pi(m-2)/m)$, which leads to $2\pi/m = 2\pi(m-2)/m$ mod $2\pi$. 
We can easily find a solution $m=3$ and substitute it back into Eq.~(\ref{eq:vector-potential}), we arrive at
\begin{align}
    \bm A(t+T/3) = \mathcal R^{-1}(2\pi/3)\bm A(t).
\end{align}
This equality indicates that the real-space path traced by the laser electric field maintains threefold rotational symmetry $\hat C_3$ and then we obtain
\begin{align}
     & \hat U_3 \hat H(t+T/3)\hat U_3^\dag = \hat H(t),            \\
     & \hat U_3 \hat{\bm J}(t+T/3)\hat U_3^\dag = \mathcal R^{-1}(2\pi/3)\hat{\bm J}(t),
\end{align}
Therefore, similarly to the case of a single CPL, we can lead to the selection rules for the HHG spectra
\begin{align}
    \bm{\mathcal J}(3\ell\Omega)=\bm 0.
    \label{eq:selection-rule}
\end{align}
We note that in the above case of two CPLs,
the dynamical symmetry associated with twofold rotation $\hat U_2$ is broken and
the relation
$\hat U_2 \hat H(t+T/2)\hat U_2^\dag = \hat H(t)$
does not hold.

The above argument based on the dynamical symmetry is always applicable regardless of the phase $\phi_i$ if we consider a continuous wave. 
However, we should note that if we consider a finite-length laser pulse, the HHG spectra can potentially change depending on the value of $\phi_i$. 
For instance, when a linearly polarized pulse with a finite $\phi_i$ is applied to graphene, the time-averaged value of the AC electric field slightly remains and it means that there is a DC field component and a resulting DC current can appear.
If the pulse length is long enough, the time-averaged DC component is negligible and then the HHG spectra are expected to obey the selection rules derived from the above argument for a continuous wave. 
This issue will be discussed later in Appendix~\ref{app:pulse-length}.

\section{Size dependence}\label{app:size}

%
\begin{figure}[tb]
    \centering
    \includegraphics[width=\linewidth]{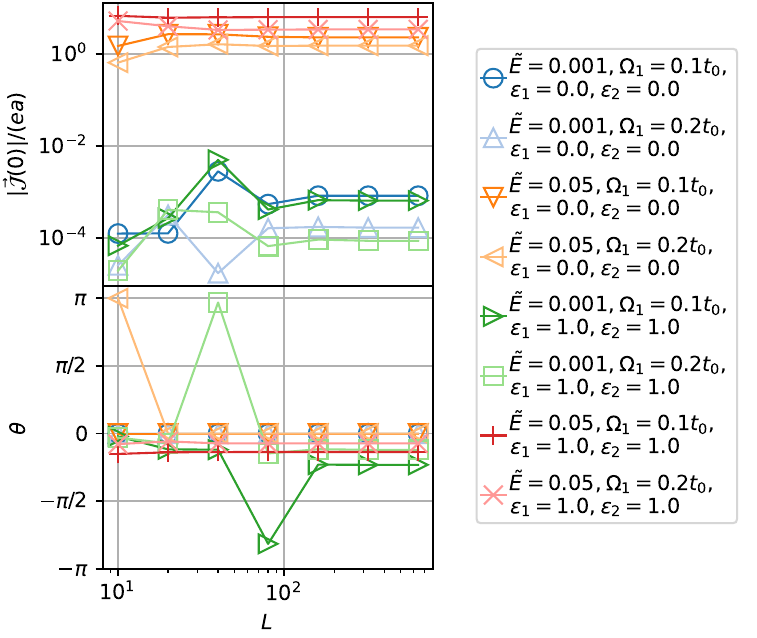}
    \caption{
    $\bk$-mesh size $L=\sqrt N$ dependence of (top row) the photocurrent magnitude and (bottom row) its direction for various parameter sets.
    We show the numerical results for $L=10,\,20,\,40,\,80,\,160,\,320,\,640$.
    We have set $\Omega_2=2\Omega_1$, $\gamma=0.05t_0$, and $\varphi=0$.
    }
    \label{fig:figureS1}
\end{figure}
%
%
Here, we briefly discuss the size dependence of the photocurrent.
Figure~\ref{fig:figureS1} shows the dependence of the photocurrent on the $\bk$-mesh size $L=\sqrt N$ for various parameter sets.
Beyond $L\geq10^2$, the photocurrent is nearly convergent, indicating that $L=160$, whose size is used in the main text, is sufficient to obtain physical quantities in the thermodynamic limit.

\section{Effect of pulse length}\label{app:pulse-length}
\begin{figure}[tb]
    \centering
    \includegraphics[width=\linewidth]{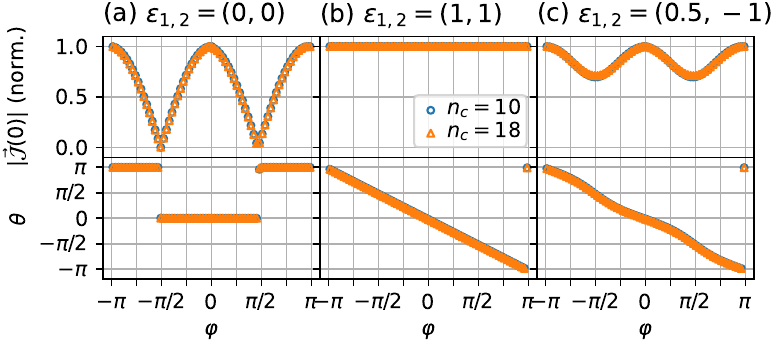}
    \caption{
    Relative phase $\varphi$ dependence of the magnitude and direction of the photocurrent in the cases of ellipticities (a) $(\epsilon_1,\epsilon_2)=(0,0)$, (b) $(1,1)$, and (c) $(0.5, -1)$.
    The magnitude of the photocurrent is normalized so that its maximum value becomes unity.
    Blue and orange dots represent the results for the systems with laser cycle numbers $n_c = 10$ and $n_c = 18$, respectively.
    Here, $n_c$ is defined as the number of cycles in the laser-pulse period $[-t_{\mathrm{FWHM}}/2, t_{\mathrm{FWHM}}/2]$.
    We have set $\Omega_1=0.2t_0$, $\Omega_2=0.4t_0$, $\gamma=0.01t_0$, and $\phi_1=0$.
    }
    \label{fig:figureS2}
\end{figure}
As we mentioned in Appendix~\ref{app:dynamical-symmetry}, there is the possibility that for a finite laser pulse,
the time averaged value of AC electric field remains finite and as a result, a DC current is induced by the finite time-averaged electric field.
To see this possibility, we compute the photocurrents in setups with different pulse lengths.

Here, we examine how the laser pulse length affects the $\varphi$ dependence of the photocurrent.
In Fig.~\ref{fig:figureS2}, we show the $\varphi$ dependence of the photocurrents in two driven systems with different laser pulse lengths $n_c$. 
Here, $n_c$ is defined as the cycle number of the laser pulse within the time period $[-t_{\mathrm{FWHM}}/2, t_{\mathrm{FWHM}}/2]$. 
The case of $n_c=10$ corresponds to the results shown in Fig.~\ref{fig:figure10}, while the case of $n_c=18$ represents a longer laser pulse. 
Comparing these two cases, we find that the photocurrents in both cases are nearly identical.
This implies that pulse length $n_c=10$ is sufficiently long for the corresponding photocurrent and HHG spectra to follow the selection rule derived by the dynamical symmetry argument, discussed in Appendix~\ref{app:dynamical-symmetry}.

\section{Photocurrent at DC limit \texorpdfstring{$\omega\to 0$}{omega -> 0} and averaged photocurrent}\label{app:average}
In the main text, we used $\bm{\mathcal J}(\omega=0)$ as the definition of photocurrent, namely, the DC limit value of the Fourier component of the current.
However, it is also possible to consider an averaged version of the photocurrent in the frequency $\omega$ space.
An averaged photocurrent may be defined as
\begin{equation}
    \bm J_{\mathrm{avg}} =\frac{2}{\Omega_1}\mathrm{Re}\qty[\int_0^{\Omega_1/2}\dd{\omega}\bm{\mathcal J}(\omega)].
\end{equation}
The upper base of the integral $\Omega_1/2$ is not important, and we may somewhat change it.
    We also consider a pseudo-averaged photocurrent as
    \begin{equation}
        \tilde{\bm J}_{\mathrm{avg}} =\mathrm{Re}\qty[\int_0^{\Omega_1/2}\dd{\omega}\bm{\mathcal J}(\omega)].
    \end{equation}
Here, we examine how this averaging operation affects the behavior of the photocurrent.
\begin{figure}[tb]
    \centering
    \includegraphics[width=\linewidth]{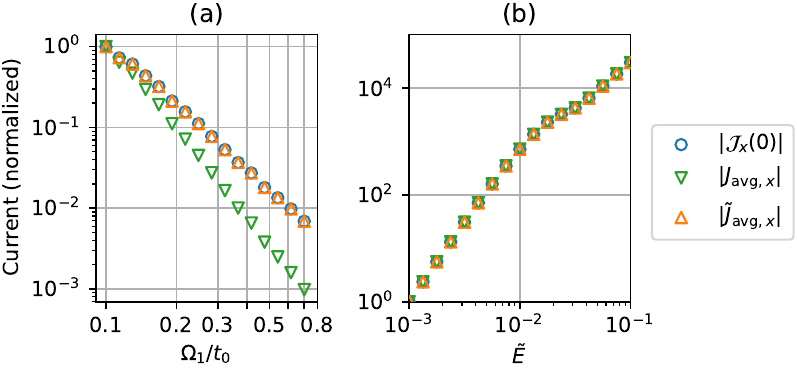}
    \caption{
    (a) $\Omega_1$ dependence of the three sorts of photocurrents $\mathcal J_x(0)$, $J_{\mathrm{avg},x}$, and $\tilde J_{\mathrm{avg},x}$ under the resonance condition of $\Omega_1=\Omega_2/2$.
    The laser intensity is set to $\tE=10^{-3}$.
    (b) $\tE$ dependence of the $x$ component of the three sorts of photocurrents under the resonance condition of $\Omega_1=\Omega_2/2 = 0.2t_0$.
    Blue, green, and orange dots in each panel represent the DC limit value $\mathcal J_x(0)$, its averaged value $J_{\mathrm{avg},x}$ and pseudo averaged one $\tilde J_{\mathrm{avg},x}$, respectively.
    Their values are normalized such that the maximum value becomes unity when (a) $\Omega_1=0.1 t_0$ and (b) $\tilde E=10^{-3}$.
    The other parameters are $\gamma=0.05t_0$ and $\epsilon_1=\epsilon_2=\phi_{1,2}=0$.
    }
    \label{fig:figureS3}
\end{figure}
Figure~\ref{fig:figureS3} compares the (a) $\Omega_1$ and (b) $\tE$ dependence of the photocurrent $\mathcal J_x(0)$ with the averaged versions, $J_{\mathrm{avg},x}$ and $\tilde J_{\mathrm{avg},x}$, by plotting them simultaneously.
The results show that $\mathcal J_x(0)$ and $\tilde J_{\mathrm{avg},x}$ match with high precision, aside from a constant factor.
While $J_{\mathrm{avg},x}$ also exhibits a similar behavior to $\mathcal J_x(0)$, the $\Omega_1$ dependence deviates from that of $\mathcal J_x(0)$ caused by the factor $2/\Omega_1$. 
We note that the power-law difference between $\mathcal J_x(0)$ and $J_{\mathrm{avg},x}$ stems from the trivial factor $2/\Omega_1$ and therefore one can easily estimate the power-law behavior of a sort of the photocurrent with another one. 
From these results, we conclude that the pseudo-averaging operation, which includes finite frequency contributions, does not qualitatively alter the nature of the photocurrent,
confirming that the photocurrent in the main text are sufficiently reliable as an observable.

%

\end{document}